# Results of testing for presence of satellites near 18 Melpomene and 532 Herculina by the speckle-interferometry method.

I.A. Sokova,[1] E.N. Sokov,[1,2] V.V. Dyachenko,[2] D.A. Rastegaev[2] and Yu.Yu. Balega[2]

[1]*Pulkovo Observatory of RAS, St. Petersburg, Russia;*

[2]*Special Astrophysical Observatory, Nizhnij Arkhyz, Russia;*

Abstract. In this work we present results of searching for satellites near the 18 Melpomene and 532 Herculina asteroids, which were predicted in 1978 from analysis of observations at the occultation moments of HD 47239 and HR 5584 by these asteroids respectively. In addition, we looked for satellites of the HD 47239 and HR 5584 stars. During several observational periods at the 6-m BTA telescope (SAO RAS) we did not detect any satellites near the 18 Melpomene and 532 Herculina asteroids, and also HR 5584. In February 2016 we clearly detected a satellite with $\rho \approx 0.01 \div 0.02$ arcsec close to HD 47239. Thus, the 18 Melpomene asteroid is likely to be single.

## 1. Introduction

On December 11, 1978, R. M. Williamon from Fernbank Science Center (Atlanta) carried out photometric observations of occultation of SAO 114159 (HD 47239) by the 18 Melpomene asteroid. In addition to the main occultation he detected a secondary fall of brightness. He predicted that this feature was caused by presence of a satellite of 18 Melpomene with a diameter of 37 km. Nevertheless the author predicted that this secondary occultation may be caused by a satellite close to HD 47239. He estimated the distance between components as 0.01 - 0.02 arcsec. Though no satellite of HD 47239 was detected, he predicted a satellite for 18 Melpomene.

Besides, in July 1978 the similar feature was detected at the moment of occultation of the HR 5584 star by the asteroid 532 Herculina. In this case the author estimated the diameter of a satellite as 50 km, orbiting around 532 Herculina at $\rho \approx 0.8$ arcsec.

The authors of this work have set a goal to test presence of satellites close to these asteroids and stars.

## 2. Results

Speckle-interferometric observations of the asteroids 18 Melpomene, 532 Herculina and the stars HD 47239, HR 5584 were carried out with the 6-m BTA telescope since the end of 2014.

Results of observations did not reveal any evidence for a satellite near the 18 Melpomene asteroid. However, an image of the HD 47239 star which was occulted by the 18 Melpomene asteroid in 1978 allowed us detecting a close satellite at a distance of about $0.01 \div 0.02$ arcsec from the star. This satellite is fainter than HD 47239 by about 3 st. magn. The discovered satellite of HD 47239 could be a cause of the secondary minima in occultation observations of 18 Melpomene and this star in 1978.

Thus, the goal of the searching for satellite of 18 Melpomene can be closed. Most likely, 18 Melpomene is a single asteroid.

The speckle-interferometric observations of the 532 Herculina and the HR 5584 stars show no signs of satellites, though the size and distance from the main body of the satellite predicted by the authors of Taylor, Dunham (1978) are within capability of the speckle-interferometer at the 6-m BTA telescope. Thus, the question about the cause of the observed secondary minima at the moment of occultation of HR 5584 by 532 Herculina in 1978 is open and it demands further investigations.

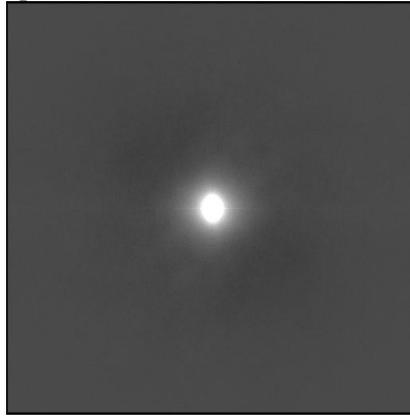

Figure 1.    The BTA image of 18 Melpomene asteroid (February, 2016).

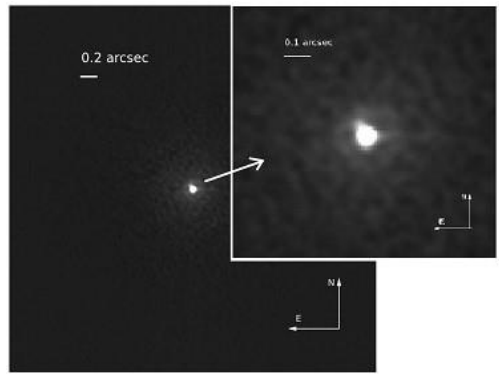

Figure 2. The BTA image of the HD 47239 star with a detected satellite (February 20, 2016).

## 3.    Acknowledgements

This work was made with support by the Russian Foundation for Basic Research (project No. 16-02-01183) and the Russian Science Foundation grant No. 14-50-00043.

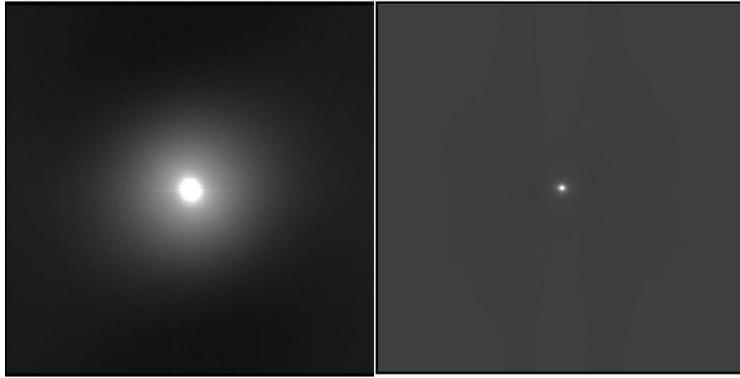

Figure 3. The BTA image of the 532 Herculina asteroid (left) and HR 5584 (right) (September, 2016).